\documentstyle[aps,prl,epsfig]{revtex}
\begin{document}
\draft
\twocolumn[\hsize\textwidth\columnwidth\hsize\csname %
@twocolumnfalse\endcsname

\title{From local to nonlocal Fermi liquid in doped antiferromagnets}
\author{P.~Prelov\v sek$^{1,2}$, J.~Jakli\v c$^{1,*}$, and K.~Bedell$^3$}
\address{ $^{1}$J.~Stefan Institute, 1000 Ljubljana, Slovenia }
\address{ $^{2}$Faculty of Mathematics and Physics, University of
Ljubljana, 1000 Ljubljana, Slovenia }
\address{$^{3}$Boston College, Chestnut Hill, MA 02467}
\date{\today}
\maketitle
\begin{abstract}\widetext
The variation of single-particle spectral functions with doping is
studied numerically within the $t$-$J$ model. It is shown that
corresponding self energies change from local ones at the 
intermediate doping to strongly nonlocal ones for a weakly doped
antiferromagnet. The nonlocality shows up most clearly in the
pseudogap emerging in the density of states, due to the onset
of short-range antiferromagnetic correlations.

\end{abstract}
 ]
\narrowtext

The understanding of electron spectral properties in the normal-state
of cuprate superconductors remains one of the hardest challenges for
theoreticians. The single-electron spectral function is directly
related to angle-resolved photoemission spectroscopy (ARPES)
experiments, which made impressive progress in the last decade
\cite{shen}, in particular in the studies of the development as
function of the hole-doping \cite{mars,ding} with respect to the
reference antiferromagnetic (AFM) material. ARPES on optimally doped
materials reveals electronic excitations with a large Fermi surface
(FS) with the volume consistent with the normal Fermi liquid (FL)
picture. The shape of spectral functions is more controversial
\cite{shen}, being possibly closer to the concept of a marginal FL
\cite{varm} than that of a usual Landau FL. Recently most interesting
ARPES results have been obtained for underdoped Bi- SrCaCuO compounds
\cite{mars,ding}, where a leading-edge shift at $T>T_c$ indicates a
feature consistent with $d$-wave superconductivity persisting in the
normal state. A pseudogap (although much larger) is observed also in
the integrated photoemission of LaSrCuO cuprates \cite{ino}. These
phenomena are closely related to relevant energy scales in doped
AFM. In underdoped cuprates experiments confirm that besides $T_c$ a
crossover $T^*>T_c$ \cite{batl}, pronounced in the spin susceptibility
$\chi_0$ and in transport quantities $\rho(T)$, $R_h(T)$, and a lower
one $T_{sg}$, most clearly seen in the NMR relaxation.

Most theoretical calculations of spectral functions 
$A({\bf k},\omega)$ in doped 2D AFM rely on numerical approaches
\cite{dago}. The exact diagonalization results for the $t$-$J$ model
\cite{step} and the quantum Monte Carlo results for the Hubbard model
\cite{bulu} provide the evidence that prototype models indeed yield a
large FS at the intermediate doping with the quasiparticle (QP)
dispersion close to the experiment and a 2D tight-binding band. A
recent application of the finite-temperature Lanczos method
\cite{jplanc,jprev} in the same regime allows for additional
conclusions for spectral functions \cite{jpspec}:
a) $A({\bf k},\omega)$ is strongly asymmetric with respect to the FS,
with underdamped QP for $k>k_F$ and essentially overdamped QP for
$k<k_F$, c) the self-energy follows the marginal FL form
${\rm Im}\,\Sigma\sim\gamma(|\omega|+\xi T)$. The regime of low but
finite doping is even harder to examine. Among the few studies of
$A({\bf k},\omega)$ only the quantum Monte Carlo results \cite{preu}
seem to indicate the existence of a pseudogap.

In the present Letter we point out two new and unexpected features of
the $t$-$J$ model as a function of doping. The first is that at
intermediate doping $A({\bf k},\omega)$ is well accounted for by the
assumption of the local FL \cite{enge}, where the self-energy is like
that recently invoked in the context of infinite-dimensional models
\cite{geor}. The second new feature is in the underdoped
regime. Reducing the doping towards an undoped AFM the nonlocality
becomes essential, leading to a pseudogap in the single-electron
density of states (DOS).

We study the planar $t$-$J$ model as the prototype model for strongly
correlated electrons in a doped AFM,
\begin{equation} 
H=-t\sum_{\langle ij\rangle s}(\tilde{c}^\dagger_{js}\tilde{c}_{is}+
\text{H.c.})+J\sum_{\langle ij\rangle} ({\bf S}_i\cdot {\bf S}_j -
{1\over 4} n_i n_j) \label{eq1}
\end{equation}
with projected operators $\tilde{c}_{is}=c_{is}(1-n_{i,-s})$
preventing the double occupation of sites. The Green's function can
be defined in the usual way $G({\bf k},\omega)=\langle\langle
\tilde{c}_{{\bf k}s};\tilde{c}^\dagger_{{\bf k}s}
\rangle\rangle_{\omega+\mu}$ where $\mu$ is the chemical potential.

It should be stressed that $A({\bf k},\omega)=-(1/\pi){\rm Im}G({\bf
k},\omega)$ is not normalized to unity \cite{step,jpspec}, since
$\langle \{\tilde c_{{\bf k} s},\tilde c^{\dagger}_{{\bf k} s}\}_+
\rangle = (1+c_h)/2=\alpha<1$. This introduces some ambiguity in the
definition of $\Sigma$. We choose here the one which guarantees the
analytical behavior, $\Sigma({\bf k},\omega \to \pm\infty)=0$
\cite{prel} (in contrast to the definition in Ref.\cite{jpspec}),
\begin{equation}
G({\bf k},\omega)= {\alpha\over \omega +\mu -\zeta_{\bf k} -
\Sigma({\bf k},\omega) },\label{eq2} 
\end{equation}
The 'free' propagation term $\zeta_{\bf k}$ is nontrivial, since the
hopping term in Eq.(\ref{eq1}) involves projected operators. It is
given by \cite{prel}
\begin{eqnarray}
\zeta_{\bf k}&=& {1\over \alpha} \langle \{[\tilde c_{{\bf k} s},H],
\tilde c^{\dagger}_{{\bf k} s}\}_+\rangle= \bar {\epsilon} +
\epsilon_{\bf k}, \nonumber \\
\epsilon_{\bf k} &=& \eta t \gamma_{\bf
k}, \qquad \eta= \alpha + {1\over \alpha} \langle S_i^z S_j^z \rangle,
\label{eq3}
\end{eqnarray}
where $\gamma_{\bf k}= \sum_{j~ n.n. 0} {\rm exp}(i{\bf k} \cdot {\bf
r}_j)$. Note that $\eta$ represents the overall renormalization of the
spectral dispersion, not just of the QP part. Since n.n.\ correlation
$\langle S_i^z S_j^z \rangle$ decreases with doping one can conclude
from Eq.(\ref{eq3}) that at the intermediate doping $\eta \sim \alpha
\alt 0.5$ while in a weakly doped AFM $\eta \sim 0.27$.

We focus on the doping dependence of the $A({\bf k},\omega)$, the
corresponding $\Sigma({\bf k},\omega)$ and the DOS, defined by ${\cal
N}(\epsilon)=(2/N)\sum_{{\bf k}} A({\bf k}, \epsilon-\mu)$.
Calculations are performed at $T>0$ and on small square lattices with
$N=16, 18$ and $20$ sites using the finite-$T$ Lanczos method
\cite{jplanc,jprev}. We fix $J=0.3~t$ to be representative of strongly
correlated systems and of cuprates in particular, where $t\sim
0.4$~eV. It is important to realize that the results are meaningful
for $T>T_{fs}$. $T_{fs}$ is the temperature where finite-size effects
start to dominate. In available systems $T_{fs}\agt 0.1~t$ at the
intermediate doping $0.1<c_h<0.25$ while it is increasing towards
undoped AFM, e.g.\ $T_{fs}\agt 0.2~t$ for $c_h \to 0$ \cite{jprev}.

Results for $A({\bf k},\omega)$ at intermediate doping have been
partly presented and discussed previously \cite{jpspec}. We point out
here a novel aspect of the results, i.e.\ a nearly ${\bf k}$
independent self-energy, $\Sigma({\bf k},\omega) \sim
\Sigma_L(\omega)$. We show in Fig.~1a Re$\Sigma$ obtained for
$c_h=3/16$, which reveal a weak variation with ${\bf k}$. This
behavior is characterstic of a local FL \cite{enge,geor}. This is in a
sharp contrast to a strong ${\bf k}$ dependence of the self-energy in
the underdoped case in Fig.~1b. One of the essential consequences of
the local FL is that the DOS at the FS ${\cal N}(\mu)$ is independent
of the self-energy, being (usually) equal to that of noninteracting
fermions. Note, however, that in the $t$-$J$ model also the 'free'
electron dispersion $\zeta_{\bf k}$ is renormalized, as follows from
Eq.(\ref{eq3}). Nevertheless the DOS for $c_h =3/16, 4/16$
\cite{jpspec} reveal a smooth variation ${\cal N}(\epsilon \sim
\mu)\propto 1/\alpha$.
\begin{figure}[htb]
\begin{center}
\epsfig{file=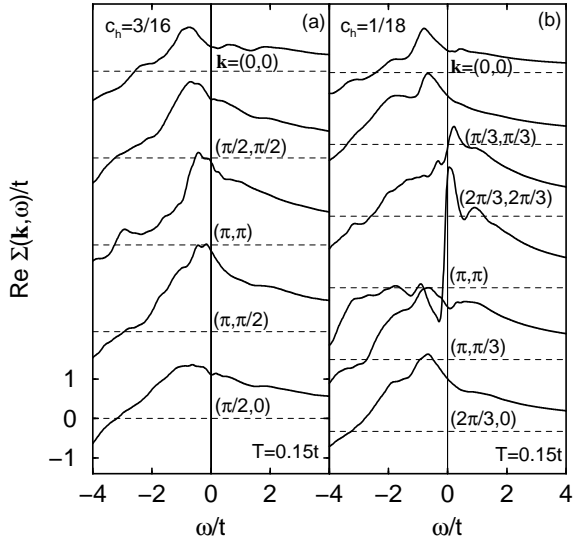,width=75mm}
\end{center}
\caption{ Self energies Re$\Sigma({\bf k},\omega)$ for the doping: a)
$c_h=3/16$, and b) $c_h=1/18$, both for $T=0.15~t$.}
\end{figure}

The local FL character is lost entering the underdoped regime. We show
in Fig.~2 $A({\bf k},\omega)$ for the lowest concentrations $c_h=0$
and $c_h \sim 0.06$, combining systems with $N=16$ and $N=18$ sites,
for the lowest $T\sim T_{fs}$. Rather well understood limit (at least
for $T \to 0$) is $c_h = 0$, where one is studying a hole in an spin
background with the long range AFM order. Note that $T>0$ and $c_h
\to 0$ requires $\mu\to -\infty$ hence it is meaningful to present
$A({\bf k},\epsilon)$. Consistent with analytical approximations
\cite{kane} the spectral function shows a coherent QP peak with 
a dispersion on the energy scale determined by $J$ (bandwidth $W \sim
2J$), with the maximum at ${\bf k}^*=(\pi/2,\pi/2)$ with $\epsilon_0
\sim 2t$. In addition to a QP feature there is a broad incoherent
background for $\epsilon \ll \epsilon_0$. The latter has still some
structure, which must be related to the AFM order, since it disappears
for $T>J$. For $c_h=0$ the dispersion corresponds to a double AFM
unit cell hence the local FL cannot apply.

\begin{figure}[htb]
\begin{center}
\epsfig{file=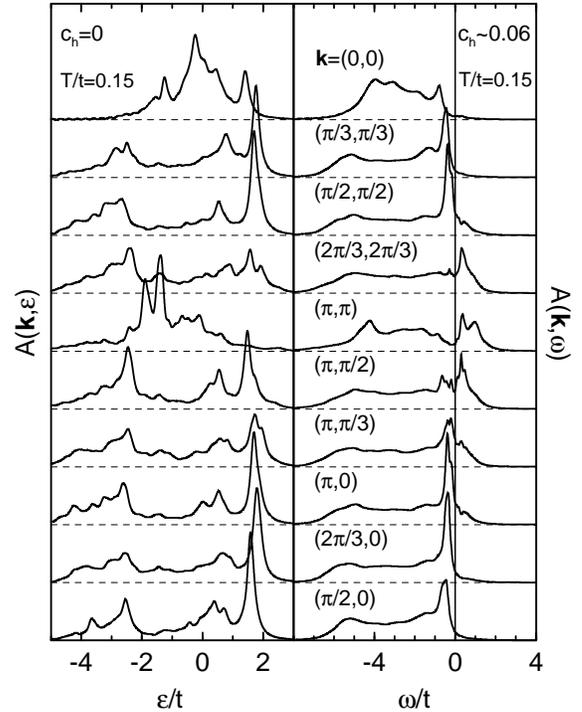,width=75mm}
\end{center}
\caption{Spectral functions for the undoped AFM (left), and for system
with finite doping $c_h=1/N$, where results for $N=16, 18$ are
combined for fixed $T=0.15~t$.}
\end{figure}

More challenging are results for $c_h \sim 0.06$, also shown in
Fig.~2. $A({\bf k},\omega)$ in this regime merge some properties of
the undoped AFM and of the intermediate doping. As expected for
$c_h>0$ there is besides a hole ($\omega<0$) part also an electron
part with $\omega>0$, its integrated intensity increasing as
$2c_h$. It is evident that the electron part is significant only for
$k>k_F$, i.e.\ for ${\bf k}$ outside the FS which seems to be close to
the square AFM Brillouin zone. Another fact is that in
$A(k>k_F,\omega)$ with a QP peak at $\omega>0$ there is also a shadow
feature for $\omega<0$, most pronounced for ${\bf
k}=(\pi,\pi/3),(\pi,\pi/2)$ as well as for ${\bf k}={\bf
Q}=(\pi,\pi)$. In fact the hole part shows a dispersion very analogous
to the one at $c_h=0$, since the dispersion seems to fold back for
$k>k_F$. This is best seen near the X point ${\bf k}=(\pi,0)$ where
the hole dispersion does not seem to reach $\omega=0$, hence an
apparent gap between the hole and the electron QP persists. The
folding effect is hardly visible along the diagonal $\Gamma$-$M$,
i.e.\ for ${\bf k}=(x,x)$, where the QP peak disperses through the FS
in a more normal way.

Gap features appear best visible in the DOS which as a local quantity
is less affected by finite size effects. In Fig.~3 we show $\cal
N(\epsilon)$ for three lowest dopings $c_h=0/20, 1/20, 2/18$ and
different $T\leq J$. For $c_h>0$ the vertical line denotes the
chemical potential $\mu(T=0.1~t)$. While at larger doping the DOS does
not show any sign of a gap at the the Fermi energy $\epsilon \sim
\mu$ (at least not for $T\agt T_{fs}$) \cite{jpspec}, it is evident
from Fig.~3 that the pseudogap starts to emerge for $c_h=2/18$ and is well
pronounced for $c_h=1/20$.

\begin{figure}[htb]
\begin{center}
\epsfig{file=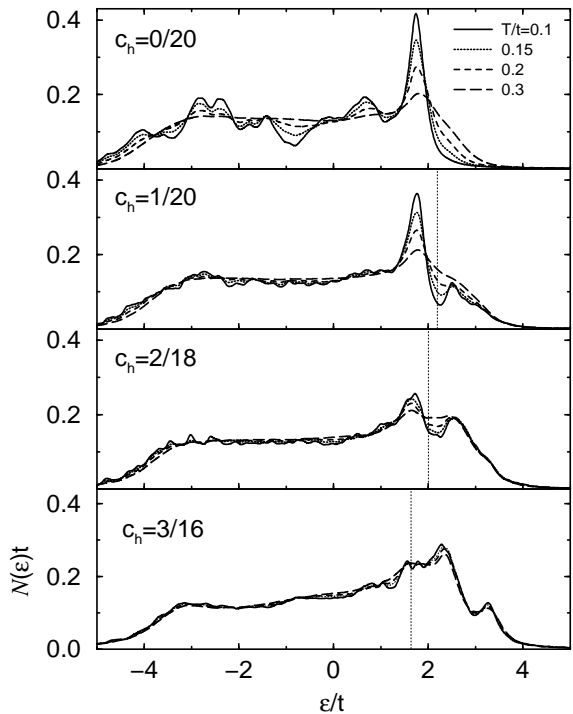,width=75mm}
\end{center}
\caption{Density of states ${\cal N(\varepsilon)}$ for systems with
various doping $c_h=0/18,1/18,2/18$ and different $T$. The thin
vertical lines denote the chemical potemtial $\mu(T=0.1~t)$.}
\end{figure}

The temperature dependence of the DOS is very instructive. It is
plausible that for $c_h=0$ the QP (spin-polaron) propagation
\cite{kane} becomes incoherent for $T>J$ due to lost AFM short range
order, hence a QP peak in $\cal N(\epsilon)$ broadens. It is
interesting to note that at $T\sim J$ some sharper resonances as well
disappear in the 'incoherent' part $\epsilon\ll \epsilon_0$. These
resonant features at $c_h=0$ and $T\to 0$ can be interpreted as real
excited resonance states of the spin-polaron \cite{kane}. The latter
are smeared out due to the spin disorder at $T>J$ or due to the
presence of other holes as concluded from Fig.~3 for $c_h>0$ where no
such resonances are visible.

A novel feature is the pseudogap at $\epsilon\sim\mu$. The latter is
not a finite-size effect since very similar results are obtained e.g.\
for systems $1/16$ and $1/18$ \cite{jpspec}. The structure disappears
for $T>J$, hence it must be related to the onset of the short-range
AFM order. Also its energy span $\Delta \epsilon <t$ seems to be
determined rather by $J$. The pseudogap is inconsistent with the
concept of the local FL, since the latter assumption would lead to an
enhanced $\cal N(\mu)$ for $c_h \to 0$ due to reduced $\eta$,
Eq.(\ref{eq3}). Hence the pseudogap can be accounted only by a
pronounced nonlocality, i.e.\ ${\bf k}$ dependence of $\Sigma({\bf
k}\sim {\bf k}_F, \omega \sim 0)$.

The underlying scenario is that the system remains a FL with a well
defined FS and a conserved (large) FS volume. This requires at least
Im$\Sigma({\bf k}_F,0) \to 0$ for $T\to 0$ whereby the FS is
determined by $\zeta_{{\bf k}_F}=\mu-{\rm Re}\Sigma({\bf k}_F,0)$. Our
numerical results for a weakly doped AFM seem to be qualitatively
consistent with these assumptions. Note that alternative scenarios
have been proposed for the regime of weak doping e.g.\ invoking a
small hole-pocket FS \cite{eder}. The FL assumption leads to the usual
form for the DOS,
\begin{equation}
{\cal N}(\mu)={2\alpha \over N}\sum_{{\bf k}} \delta(\mu-\zeta_{{\bf
k}} - {\rm Re}\Sigma({\bf k},0) ).\label{eq4}
\end{equation}
The pseudogap thus requires a large $|d{\rm Re}\,\Sigma({\bf k},0)/d
{\bf k}|$ at ${\bf k}={\bf k}_F$. In Fig.~1b we show ${\rm
Re}\,\Sigma({\bf k},\omega)$ at $c_h=1/18$ and fixed
$T=~0.15~t$. First we realize that Re$\Sigma({\bf
k},0)=\mu-\zeta_{{\bf k}}$ locates the presumed FS again close to the
AFM zone boundary. The differences with respect to intermediate doping
\cite{jpspec} are easily observed. While for ${\bf k}$ inside the FS
the self-energy is quite similar to the local FL, there is an
increasing deviation from the latter for $k>k_F$. In particular this
shows up as an oscillation in Re$\Sigma({\bf k},\omega)$ most
pronounced for ${\bf k}={\bf Q}$. The effect is so strong that it
leads for most $k>k_F$ to a double solution for a QP peak, i.e.\
$E_{{\bf k}}=\zeta_{{\bf k}}+{\rm Re}\Sigma({\bf k},E_{\vec k})-\mu$,
explaining the shadow features and folded spectra in Fig.~2 for most
${\bf k}$ outside the FS.

The double solutions for $E_{{\bf k}}$ are analogous to those
appearing within the spin-bag theory \cite{kamp} taking into account
the AFM short-range order. The essential difference is however an
evident asymmetry since no such effect appears for ${\bf k}$ inside
the FS. To simulate the observed corrections to $\Sigma_L$ due to AFM
correlations can be designed. For a long-range AFM order one could
expect the form
\begin{equation}
\Sigma({\bf k},\omega)\sim\Sigma_L(\omega) + a({\bf k}) G( {\bf
k}+{\bf Q},\omega).\label{eq5}
\end{equation}
The expression (\ref{eq5}) naturally leads to oscillations in
$\Sigma({\bf k},\omega)$ and possible double QP solutions, in
particular near the half-filling where the FS is close to the nesting
AFM zone boundary. In order to account for the pronounced asymmetry
with respect to FS one has to assume an essential dependence of
$a({\bf k})$, i.e.\ $a(k \ll k_F) \sim 0$ and its strong increase near
$k_F$.

It is however still feasible that in a weakly doped AFM the emerging
pseudogap coexists with a well defined large FS. AFM correlations open
a pseudogap between the hole-like and electron-like QP resonances
coexisting for ${\bf k}\sim{\bf k}_F$, visible in particular near the
X point in Fig.~2. Here it seems that the QP dispersion does not
traverse the FS. But the assumption Im$\Sigma({\bf k},0)=0$ at $T\to
0$ leads to another QP crossing the FS. The latter peak must be of a
small QP weight $\tilde Z= Z_{{\bf k}_F}$. This could be consistent
with the inequality $\tilde Z<2 c_h$ \cite{jpspec} derived on the
assumption of the monotonous fall-off of $\bar n_{{\bf k}}$. Such a QP
excitation would be hard to observe in numerical calculations as well
as in experiments since it can be easily overdamped either by $T>0$ or
other effects.

Let us turn to the discussion of the relation of our results to
photoemission experiments on cuprates. The latter capture mostly the
hole-part $\omega<0$. For the intermediate doping our $A({\bf
k},\omega)$ are qualitatively consistent with ARPES measurements
\cite{shen}, as discussed in Ref.\cite{jpspec}. ARPES on underdoped
BiSrCaCuO compounds \cite{mars} shows an opening of a pseudogap near
the X point where the FS seems to disappear. On the other hand the QP
peak disperses through the FS along the $\Gamma$-M direction, rather
like in a normal FL. Both facts are well consistent with results in
Fig.~2. Recently the DOS has been measured via the integrated
photoemission (also the inverse part) for LaSrCuO material in a wide
range of doping \cite{ino}. More reliable are emission spectra related
to ${\cal N}^+(\omega)={\cal N}(\omega+\mu)f(\omega)$, giving
information for $\omega<0$. In the overdoped systems $c_h>0.2$ ${\cal
N}^+(\omega<0)$ appears flat. On the other hand, for $c_h<0.17$ a
pseudogap starts to emerge gradually, e.g.\ the inflection point in
${\cal N}^+(\omega)$ moves from $\omega \sim 0$ towards $\omega \sim
-0.2$eV at lowest doping. This is quite close to our results and
values in Fig.~3. The pseudogap scale and its doping dependence seems
thus to be related to $T^*$ \cite{batl}, appearing e.g.\ as the
maximum in $\chi_0$. Note however that such a gap is much larger than
the leading-edge shift $\sim 30$meV found in ARPES \cite{mars} in
underdoped samples.

In conclusion, we have shown that within the $t$-$J$ model the
spectral function changes with doping from that of a local FL to a
strongly nonlocal one. It seems plausible that the locality of the
self-energy at intermediate doping is related to the coupling of QP to
large (mostly local) quantum fluctuations of the spin subsystem
\cite{prel} whereby its anomalous behavior emerges due to the
frustration induced by holes. The locality has several important
consequences. Apart form the mentioned effect on ${\cal N}(\mu)$, one
key feature is the reduction of FL parameters to only two, which are
even simply related \cite{enge}. It is still a question whether such a
local FL has properties analogous to ones obtained in infinite-D
models \cite{geor}.

In contrast, the nonlocality of $\Sigma({\bf k},\omega)$ induced by
short-range AFM correlations becomes essential for the regime of weak
doping, and is dominating the low-energy behavior $\omega \sim 0$. The
most pronounced effect is the opening of the pseudogap in the
DOS. $A({\bf k},\omega)$ in this regime can reveal a coexistence of a
hole-like dispersion analogous to undoped AFM and an electron-like
resonance. One of the challenging questions remains whether in
underdoped systems we are still dealing with a well defined large FS,
whereby our results do not contradict this scenario. It should also be
pointed out that nonlocality is crucial to account for the marked
difference in the low-doping regime between nearly free-fermion-like
Wilson ratio and a strongly enhanced (possibly even singular)
compressibility \cite{jprev}, indicating the closeness to a charge
instability.

The authors acknowledge the support of the SLO-US Grant and
the DOE Grant DEFG0297ER45626.

\noindent
$^*$ Present address: Cadence Design Systems, D-85540 Haar, Germany.

%\begin{figure}
%\caption{ Self energies Re$\Sigma({\bf k},\omega)$ for the doping: a)
%$c_h=3/16$ and b) $c_h=1/18$, both for $T=0.15~t$.} \label{fig1}
%\end{figure}

%\begin{figure}
%\caption{ Spectral functions for the undoped AFM (left), and for system
%with finite doping $c_h=1/N$, where results for $N=16, 18$ are
%combined for fixed $T=0.15~t$. }\label{fig2}
%\end{figure}

%\begin{figure}
%\caption{ Density of states ${\cal N(\varepsilon)}$ for systems with
%various doping $c_h=0/18,1/18,2/18$ and different $T$. The thin
%vertical lines denote the chemical potemtial $\mu(T=0.1~t)$.}\label{fig3}
%\end{figure}

\end{document}